# Sleep Staging from Electrocardiography and Respiration with Deep Learning

Haoqi Sun, PhD[1], Wolfgang Ganglberger, MSc[1], Ezhil Panneerselvam, MD[1], Michael J. Leone, MSc[1], Syed A. Quadri, MD[1], Balaji Goparaju, MSc[1], Ryan A. Tesh, BS[1], Oluwaseun Akeju, MD[2], Robert J. Thomas, MD[3], M. Brandon Westover, MD, PhD[1]

[1] Department of Neurology, Massachusetts General Hospital, Boston, MA, USA
[2] Department of Anesthesia, Critical Care and Pain Medicine, Massachusetts General Hospital, Boston, MA, USA
[3] Division of Pulmonary, Critical Care & Sleep, Department of Medicine, Beth Israel Deaconess Medical Center, Boston, MA, USA

## Abstract

**Study Objective**: Sleep is reflected not only in the electroencephalogram but also in heart rhythms and breathing patterns. Therefore, we hypothesize that it is possible to accurately stage sleep based on the electrocardiogram (ECG) and respiratory signals.

**Methods**: Using a dataset including 8,682 polysomnographs, we develop deep neural networks to stage sleep from ECG and respiratory signals. Five deep neural networks consisting of convolutional networks and long short-term memory networks are trained to stage sleep using heart and breathing, including the timing of R peaks from ECG, abdominal and chest respiratory effort, and the combinations of these signals.

**Results**: ECG in combination with the abdominal respiratory effort achieve the best performance for staging all five sleep stages with a Cohen's kappa of 0.600 (95% confidence interval 0.599 – 0.602); and 0.762 (0.760 – 0.763) for discriminating awake vs. rapid eye movement vs. non-rapid eye movement sleep. The performance is better for young participants and for those with a low apnea-hypopnea index, while it is robust for commonly used outpatient medications.

**Conclusions**: Our results validate that ECG and respiratory effort provide substantial information about sleep stages in a large population. It opens new possibilities in sleep research and applications where electroencephalography is not readily available or may be infeasible, such as in critically ill patients.

## Introduction

Characterizing sleep has primarily relied on the analysis of the electroencephalogram (EEG), supplemented by the electrooculogram and chin electromyogram [1]. Three distinct states are readily discernable through such analysis: wake, rapid eye movement sleep (REM) and non-REM sleep (NREM) [1]. Three stages of progressive depth (N1, N2, and N3) can be differentiated in NREM [1]. EEG slow-wave oscillations (about 1 – 4Hz) dominate deeper NREM sleep, while sleep spindles (about 10 – 13 Hz) and theta wave oscillations (about 4 – 8Hz) dominate lighter NREM sleep [1].

Cortical, subcortical, and brainstem systems are highly interactive throughout sleep [2-6], and their activity couples autonomic activity with the cortical activity measurable by EEG. Examples include strong sinus arrhythmia [3], blood pressure dipping [7], and stable breathing or stable flow-limitation [8]. REM sleep is characterized by highly recognizable respiratory rate and tidal volume fluctuations, and by surges in heart rate and blood pressure [9]. Wake demonstrates dominance of low-frequency heart rate variability and large amplitude movements. These observations suggest that accurate sleep staging might be possible from non-EEG signals influenced by the autonomic nervous system, such as the ECG or respiratory signals.

An accurate non-EEG method for staging state characterization would have several advantages. For example, the ECG is recorded continuously in numerous medical and situations, especially in hospitalized patients. Wearable devices increasingly measure ECG and respiration [10-12]. Cardiorespiratory signals may be obtainable in a number of ways, including contact recordings such as Withings, ballistocardiogram [13], or non-contact radar-type applications such as EarlySense [14], and SleepScore [15], etc. On the other hand, EEG can be highly abnormal in medically ill populations, making standard analysis difficult [16].

Deep learning approaches can be used to accurately estimate sleep states. We previously showed that deep neural networks can learn to score conventional sleep stages based on EEG signals obtained during overnight PSG with an accuracy of 87.5% and a Cohen's kappa of 0.805, comparable to the performance of human sleep scoring experts [17]. Here we develop deep neural networks using ECG and/or respiratory signals to classify sleep stages. Our approach is based on convolutional neural network (CNN) in combination with long-short term memory (LSTM) recurrent neural network. It is trained on a large clinical dataset, which also accounts for patient heterogeneity, spanning a wide range of ages, medications and sleep disorders.

## Methods

### Dataset
The Partners Institutional Review Board approved retrospective analysis of polysomnograms (PSG), acquired in the Sleep Laboratory at Massachusetts General Hospital from 2009 to 2016, without requiring additional consent for use in this study. PSGs were recorded adhering to American Academy of Sleep Medicine (AASM) standards. Each PSG includes one ECG channel and two respiratory effort channels recorded from chest and abdomen belts. The sampling frequency is 200 Hz for all signals. The dataset contains three major types of sleep tests: diagnostic, full-night continuous positive airway pressure (CPAP), and split-night CPAP. PSGs were annotated in 30-second non-overlapping epochs according to AASM standards as one of the five stages: wake (W), non-REM stage 1 (N1), non-REM stage 2 (N2), non-REM stage 3 (N3), and rapid eye movement (REM). Seven sleep technicians in total annotated the dataset, with one technician per PSG.

The entire dataset includes 10,121 PSGs; 9,644 were exported successfully without time mismatch or missing sleep stage annotations. We included atrial fibrillation cases because the deep learning network is supposed to work with heterogeneous data. We excluded PSGs with fewer than 100 artifact-free 30-second epochs, resulting in 8,682 PSGs. The dataset is summarized in Table 1.

**Table 1**. Dataset Summary.

| Characteristics | Value |
|---|---|
| Number of PSGs | 8,682 |
| Number of patients | 7,208 |
| Age: year, median (IQR) | 53 (41 – 63) |
| Sex: number (percentage of all patients) | |
|     Female | 2,997 (41.6%) |
|     Male | 4,189 (58.1%) |
|     Unknown due to human error | 22 (0.3%) |
| BMI: kg/m$^2$, median (IQR) | 31 (27 – 36) |
| Type of Test: number (percentage of all patients) | |
|     Diagnostic | 3,571 (49.5%) |
|     All night CPAP | 1,751 (24.3%) |
|     PSG split night | 1,798 (24.9%) |
|     Extended EEG-sleep montage | 76 (1.1%) |
|     Bedside | 9 (0.1%) |
|     Research | 3 (0.04%) |
| Apnea-Hypopnea Index (AHI, events / hour) : number (percentage of all patients) | |
|     Normal (AHI < 5) | 2,879 (39.9%) |
|     Mild (5 ≤ AHI < 15) | 1,995 (27.7%) |
|     Moderate (15 ≤ AHI < 30) | 1,468 (20.4%) |
|     Severe (AHI ≥ 30) | 866 (12.0%) |
| Respiratory Disturbance Index (RDI, events / hour) | 15.0 (5.8 – 28.4) |
| Periodic Limb Movement Index (PLMI, events / hour) | 10.4 (3.1 – 28.3) |
| Outpatient Medication Listing, by Category | |
|     Systemic | 4523 (62.7%) |
|     Hypertension | 2755 (38.2%) |
|     Sleeping | 2187 (30.3%) |
|     Antidepressant | 1874 (26.0%) |
|     Neuroactive | 1365 (18.9%) |
|     Benzodiazepine | 1297 (18.0%) |
|     Diabetic | 802 (11.1%) |
|     RLS/PLMS | 688 (9.5%) |
|     Opiate | 548 (7.6%) |
|     Z-drug | 348 (4.8%) |
|     Stimulant | 310 (4.3%) |

**Preprocessing**

Sleep staging was done in 30-second epochs following AASM standards. However, changes in heart rhythms and respiration often occur over longer time scales. For this reason, and to provide contextual information, our deep neural networks used information extending 120 seconds on both sides of each 30-second epoch, creating a 270-second epoch (4.5min, nine 30-second epochs) centered on each 30-second epoch to be scored. The goal of the deep neural networks presented herein is to classify the sleep stage of the middle 30-second epoch using information from the 270-second epoch. This is illustrated in Figure S1 in the supplementary material.

When using ECG as the input, we identified 270-second epochs with amplitude larger than 6mV or standard deviation smaller than 5μV, as they are not physiologically possible. In each 270-second epoch we extracted timings of R peaks [18] and converted the ECG to a binary sequence, where R peaks are indicated by a "1" and all other points indicated by "0". The 270-second epochs with spurious R peaks were identified using the ADARRI [19] method. 270-second epochs with less than 20 R peaks per minute were also identified, as they are not physiologically possible. About 25% of the 270-second epochs were identified as artifact. In total, there were 5,964,359 270-second epochs.

When using chest and abdominal respiratory effort as the input, 270-second epochs with amplitude larger than 6mV or standard deviation smaller than 10μV were identified. Respiratory signals were down-sampled to 10Hz. About 10% of the 270-second epochs were identified as artifact. In total, there were 6,847,246 270-second epochs for the chest signal; and 6,749,286 270-second epochs for the abdominal signal.

When using pairs of signals as the input, the 270-second epochs where any signal modality that meet the above criteria are identified as artifact. In any of above cases, the artifactual 270-second epochs were removed when training CNN, and remained when training LSTM to preserve the temporal context.

**Deep Network Architecture**

We trained five deep neural networks based on the following input signals and their combinations: 1) ECG; 2) CHEST (chest respiratory effort); 3) ABD (abdominal respiratory effort); 4) ECG+CHEST; and 5) ECG+ABD. Each deep neural network contained a feed-forward convolutional neural network (CNN) which learned features pertaining to each epoch, and a recurrent neural network (RNN), in this case long-short term memory (LSTM), to learn temporal patterns among consecutive epochs.

The CNN of the network is similar to that in Hannun et al. [20]. As shown in Figure 1A and Figure 1B, the network for a single type of input signal, i.e. ECG, CHEST or ABD, consists of a convolutional layer, several residual blocks and a final output block. For a network with both ECG and CHEST/ABD as input signals (Figure 1C), we first fixed the weights of the layers up to the $9^{th}$ residual block (gray) for the ECG network and similarly fixed up to the $5^{th}$ residual block (gray) for the CHEST/ABD network, concatenated the outputs, and then fed this concatenation into a subnetwork containing five residual blocks and a final output block. The numbers of fixed layers were chosen so that the outputs of layers from different modalities have the same shape (after padding zeros), and were then concatenated.

The LSTM of the network has the same structure for different input signals. It is a bi-directional LSTM, where the context cells from the forward and backward directions are concatenated. For the network with ECG as input, the LSTM has two layers with 20 hidden nodes in each layer. For CHEST and ECG+CHEST, the LSTM has three layers with 100 hidden nodes in each layer. For ABD and ECG+ABD, the LSTM has two layers with 100 hidden nodes in each layer. The number of LSTM layers, number of hidden nodes, and dropout rate were determined by the method described in the next subsection.

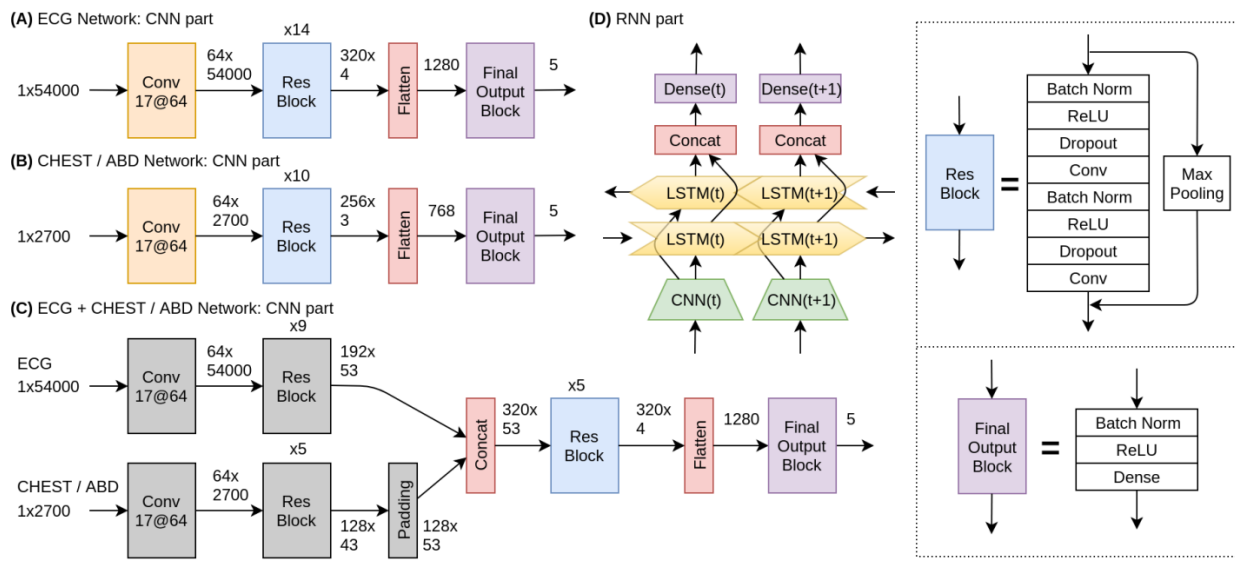

**Figure 1**. Deep neural network architecture. **(A and B)** CNN architecture using ECG, or CHEST or ABD as input. The numbers between blocks are the shapes of the output for each input 270-second epoch. For example, "320×4" means 320 channels and four time points. "17@64" in the convolution layers means kernel size 17 points and 64 kernels. The repetition number of the residual blocks (Res Block) is marked above each block. Arrows indicate the flow of network activations. **(C)** The CNN architecture when using multiple signals as input. Gray blocks mean their weights are obtained from network trained in (A) and (B), then fixed during training the network. **(D)** RNN architecture, which uses the output from the CNN from every 270-second epoch (corresponding to a 30-second epoch). The output is fed into a bidirectional LSTM, followed by concatenation of the activations from both directions, and finally into a dense layer. The legends on the right show the detailed structure of the residual block and final output block. Inside each residual block, the first convolution layer subsamples the input by 4 (stride = 4) and the max pooling skip-layer connection also subsamples the input by 4.

**Training and Evaluating the Network**
We randomly split the PSGs into a training set of 6,682 PSGs, a validation set of 1,000 PSGs and a testing set of 1,000 PSGs. Due to the large amount of data, we expect the random split should give similar distribution of the variables across these sets. We first trained the CNN, then the LSTM using the outputs from the CNN. The objective function of both CNN and LSTM is cross-entropy, a measure of the distance between two categorical distributions for classification. The networks were trained with a mini-batch size of 32, maximum epochs of 10, and learning rate 0.001 (as commonly used in deep learning). The LSTM was trained using sequences of 20 epochs (10min). We set the number of LSTM layers, number of hidden nodes, and the dropout rate as the combination that minimizes the objective function on the validation set.

Some sleep stages occur more frequently than others. For example, people spend about 50% of sleep in N2 and 20% in N3. To prevent the network from simply learning to report the dominant stage, we weighed each 270-second input signal in the objective function by the inverse of the number of epochs in each sleep stage within the training set.

For the performance metrics on the testing set, we used confusion matrices and Cohen's kappa. We show performance for staging five sleep stages according to the AASM standards (W, N1, N2, N3, R), and we additionally collapse these stages into 3 sleep super-stages, in two different ways. The first set of super-stages is "awake or drowsy" (W+N1) vs. "sleep" (N2+N3) vs. "REM sleep" (R), and The second set of super-stages is "awake" (W) vs. "NREM sleep" (N1+N2+N3) vs. "REM sleep" (R). We obtained 95% confidence intervals for kappa values by bootstrapping (sample with replacement) 1,000 times.

## Results

**Overall Staging Performance**

In Figure 2, we show the confusion matrices for predicting all five sleep stages with different input signals. Using both ECG and ABD as input signals yields the best prediction results on the testing set. This network is correct in 81.8% of wake, 55.8% of N1, 66.3% of N2, 66.6% of N3 and 92.2% of REM epochs. Most misclassifications are found between W vs. N1, N1 vs. N2, and N2 vs. N3. For example, 31% of N3 epochs are misclassified as N2, and 34% of N1 epochs are misclassified as either W or N2. This limitation is reduced when grouping the sleep stages as in Figure 3 and Figure 4, so that both epochs of REM and NREM can be classified correctly with greater than 80% accuracy.

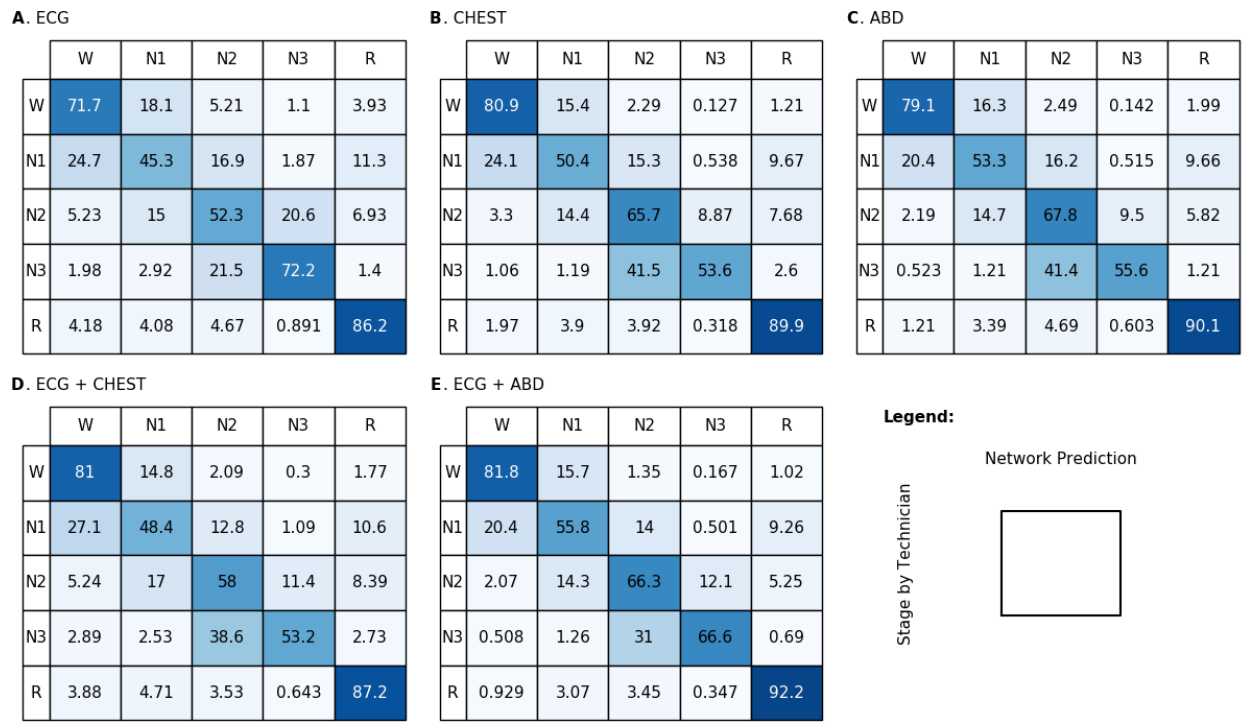

**Figure 2**. Five-stage classification confusion matrices, comparing staging by sleep technicians vs. network predictions on the 1000-PSG testing set for different input signals. Each row in the confusion matrix is the sleep stage annotated by the technician, while each column is the network prediction. The numbers are percentages.

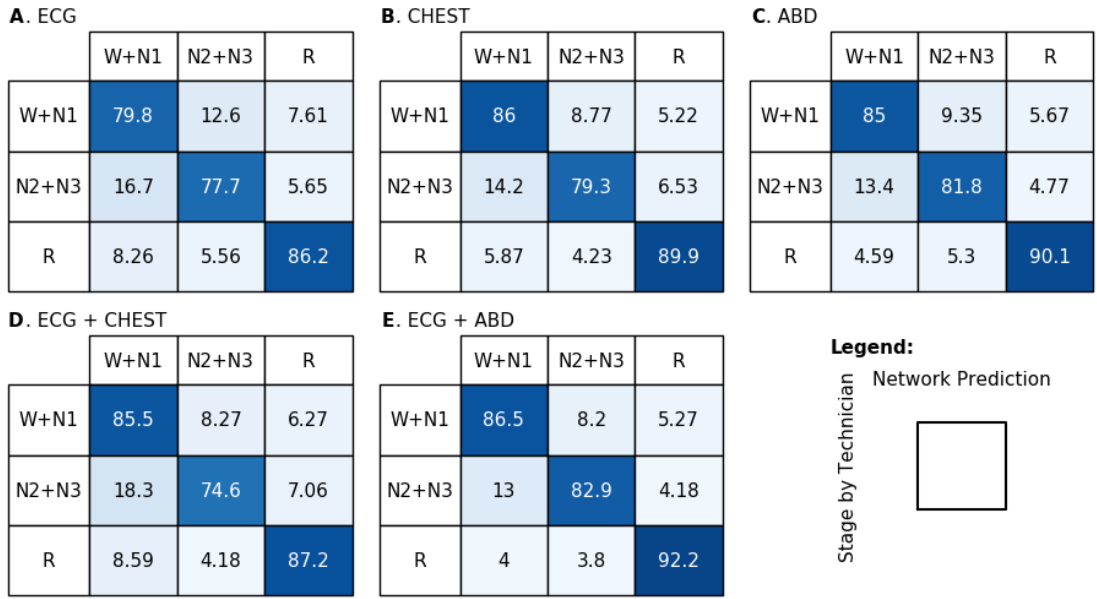

**Figure 3**. Three-stage classification confusion matrices, comparing staging by sleep technicians vs. network predictions on the 1000-PSG testing set for different input signals. The 3 "super-stages" here are: "awake or drowsy" (W+N1) vs. "sleep" (N2+N3) vs. "REM sleep" (R).

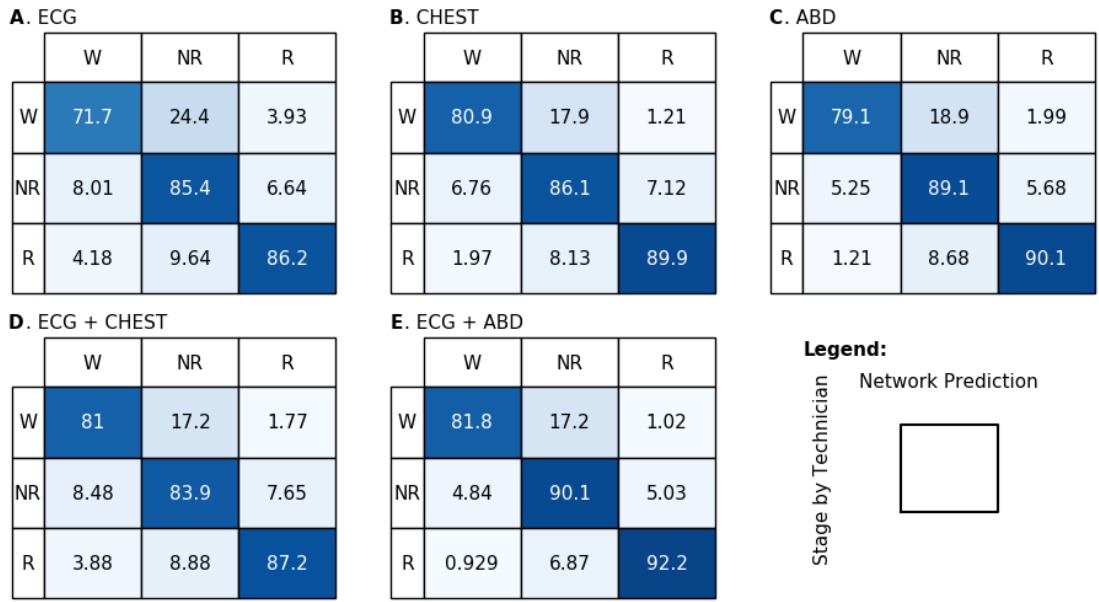

**Figure 4**. Three-stage classification confusion matrices, comparing staging by sleep technicians vs. network predictions on the 1000-PSG testing set for different input signals. The 3 "super-stages" here are: "awake" (W) vs. "NREM sleep" (N1+N2+N3) vs. "REM sleep" (R).

For every choice of input signal, we calculated Cohen's kappa, a statistic for assessing inter-reader agreement, and the corresponding 95% confidence intervals. These are shown in Table 2. ECG+ABD has the highest kappa, with values of 0.6 (all five stages), 0.74 (W+N1 vs. N2+N3 vs. REM) and 0.762 (Wake vs. NR vs. REM). Since the testing set has 1,000 PSGs ($6.6 \times 10^5$ 30-second epochs), the confidence interval is narrow. Therefore the differences between kappa values are all significant at 0.05 level.

**Table 2**. Cohen's kappa on the 1000-PSG testing set using different input signals.

| Input Signal | 5 Stages | 3 Stages | |
|---|---|---|---|
| | | W+N1 vs. N2+N3 vs. R | W vs. NR vs. R |
| ECG | 0.494 (0.492 – 0.495) | 0.649 (0.647 – 0.65) | 0.649 (0.647 – 0.651) |
| CHEST | 0.565 (0.563 – 0.566) | 0.708 (0.707 – 0.709) | 0.707 (0.706 – 0.709) |
| ABD | 0.579 (0.578 – 0.581) | 0.725 (0.723 – 0.726) | 0.738 (0.736 – 0.739) |
| ECG + CHEST | 0.506 (0.504 – 0.507) | 0.648 (0.646 – 0.65) | 0.661 (0.659 – 0.663) |
| ECG + ABD | 0.600 (0.599 – 0.602) | 0.740 (0.738 – 0.741) | 0.762 (0.76 – 0.763) |

**Staging Performance on Different Groups of Participants**

In Table 3, we show Cohen's kappa for different population groups in the testing set using ECG+ABD as the input signals. Kappa is lower in the elderly (≥ 60 years) and in people with higher Apnea-Hypopnea Index (AHI), compared to the respective control groups. Split-night studies have lower kappa values than diagnostic or CPAP nights due to different patterns before and after applying CPAP. Note again that due to the large number of epochs in the testing set, confidence intervals are narrow, and all comparisons are statistically significant at 0.05 level. The Cohen's kappa for different population groups in the testing set using other input signals are shown in Table E1-E4 in the supplementary material.

While performance is reduced with increasing AHI, the network still achieves Cohen's kappa of 0.574 for five stages; and more than 0.7 for three super-stages for severe apnea. We interpret this to mean that either autonomic features characteristic of stages are independent of sleep apnea, or more likely, that the network has learned normal, apneic, and other pathological patterns of the respiration signals change according to sleep stage. For example, REM and NREM interruptions in breathing may have distinct distributions of features such as event duration.

**Table 3**. Cohen's kappa in different groups in the 1000-PSG testing set using ECG + ABD as input.

| Category | Group | 5 stages | 3 stages | |
|---|---|---|---|---|
| | | | W+N1 vs. N2+N3 vs. R | W vs. NR vs. R |
| Age | Young: 18 ≤ Age < 40 | 0.622 | 0.760 | 0.773 |
| | Middle: 40 ≤ Age < 60 | 0.598 | 0.736 | 0.761 |
| | Old: Age ≥ 60 | 0.576 | 0.713 | 0.742 |
| Sex | Male | 0.594 | 0.728 | 0.755 |
| | Female | 0.602 | 0.746 | 0.763 |
| BMI (kg/m$^2$) | Normal: 18.5 ≤ BMI < 25 | 0.613 | 0.737 | 0.752 |
| | Overweight: BMI ≥ 25 | 0.596 | 0.736 | 0.760 |
| Type of Test | Diagnostic | 0.600 | 0.738 | 0.756 |
| | All Night CPAP | 0.608 | 0.758 | 0.770 |

|  | Split Night | 0.584 | 0.707 | 0.752 |
|---|---|---|---|---|
| AHI (per hour) | Normal: AHI < 5 | 0.608 | 0.760 | 0.773 |
|  | Mild: 5 ≤ AHI < 15 | 0.604 | 0.746 | 0.750 |
|  | Moderate: 15 ≤ AHI < 30 | 0.600 | 0.733 | 0.752 |
|  | Severe: AHI ≥ 30 | 0.574 | 0.717 | 0.742 |
| Periodic limb movement (per hour) | Normal: PLM < 5 | 0.606 | 0.736 | 0.765 |
|  | Mild: 5 ≤ PLM < 15 | 0.600 | 0.746 | 0.767 |
|  | Moderate: 15 ≤ PLM < 30 | 0.600 | 0.733 | 0.752 |
|  | Severe: PLM ≥ 30 | 0.574 | 0.717 | 0.742 |
| Medication | Antidepressant | 0.588 | 0.736 | 0.756 |
|  | Benzodiazepine | 0.602 | 0.750 | 0.761 |
|  | Diabetic | 0.589 | 0.757 | 0.770 |
|  | Herbal | 0.596 | 0.745 | 0.743 |
|  | Hypertension | 0.595 | 0.732 | 0.758 |
|  | Neuroleptic | 0.588 | 0.716 | 0.749 |
|  | Opiate | 0.600 | 0.721 | 0.766 |
|  | Neuroactive | 0.598 | 0.736 | 0.762 |
|  | Systemic | 0.599 | 0.740 | 0.766 |
|  | RLS/PLMS | 0.608 | 0.755 | 0.773 |
|  | Sleeping | 0.610 | 0.750 | 0.763 |
|  | Stimulant | 0.603 | 0.740 | 0.774 |
|  | Z-drug | 0.627 | 0.765 | 0.774 |

**Staging Performance on Individual PSGs**

In Figure 5, we show the histogram of Cohen's kappa of each individual PSG using both ECG and ABD as input. The results indicate a fair amount of heterogeneity between PSGs, where the lowest extreme has negative kappa values around -0.1 and the highest extreme has kappa values around 0.9. In Figure S2 in the supplementary material, we show the Cohen's kappa of each individual PSG using all signal types.

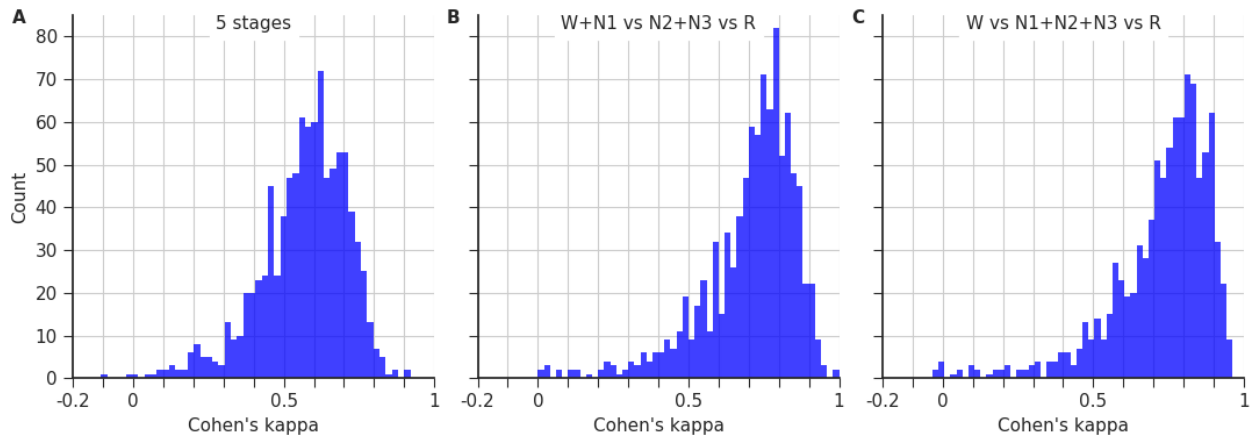

**Figure 5**. Histogram of Cohen's kappa values for individual PSGs using both ECG and ABD as input. The distributions are right-skewed.

**Dependence on temporal precision of R-peak timing in the ECG**

In face of signal noise, the deep learning network should learn robust patterns of the ECG R peak time series. To validate its robustness to signal noise, we simulated noise that preserves the mean but corrupts higher order pattern of the ECG R peaks. In Figure S3 of the supplementary material, we can see that adding zero-mean Gaussian jitter to the R peaks causes performance to drop progressively as the standard deviation of the jitter increases.

**Signal Examples**

To gain some insight into the differences in breathing and heart rhythms that the deep neural network is using to distinguish sleep stages, we show some example whole night recordings from the 1000-PSG testing set in Figure 6, Figure 7, and Figure 8. These examples are selected as "typical", meaning that they have the closest Cohen's kappa compared to the overall kappa across the testing set. The 60-second signal examples in Panel C are the signals where the deep neural network assigns the highest probability to the correct sleep stage within the recording. We can see a visible correspondence between the spectrogram and the sleep stages, as well as the mismatch between the spectrogram and EEG-based sleep stage. For example, in Figure 8, around 2 hours and 4.5 hours, the spectrogram of heart rate variability shows loss of very low frequency power, which is classified by the network as N3, but the EEG-based sleep stages contain both N2 and N3. More illustrations of the trained deep neural networks are shown in Figure S4-S12 of the supplementary material.

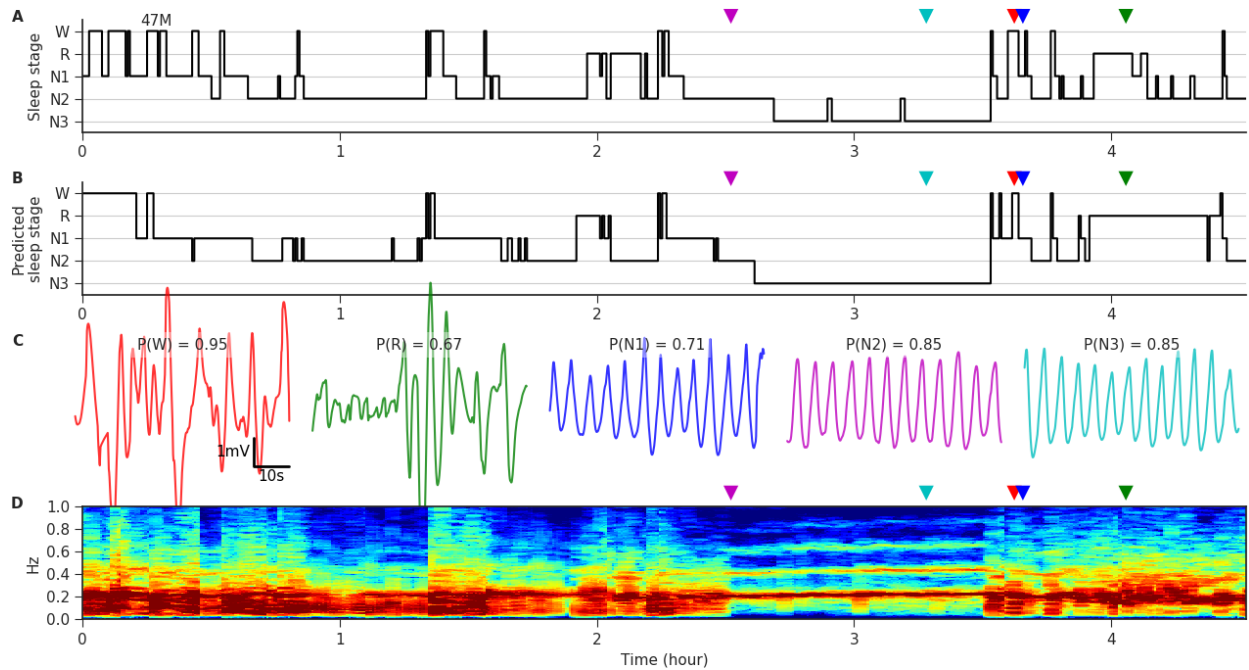

**Figure 6**. An example 47-year male. **(A)** The sleep stages over the whole night annotated by the technician (hypnogram). **(B)** The predicted sleep stages from the deep neural network using ABD respiration as input. **(C)** Example 60-second ABD segment from each sleep stage which is correctly classified and has the highest predicted probability of that stage. Different colors correspond to the triangle markers on other panels, which indicate the location of the example in the whole night recording. The number above each example signal indicates the probability of being that stage as predicted by the deep learning network. **(D)** The spectrogram of the ABD respiratory signal. The y-axis indicates the frequency.

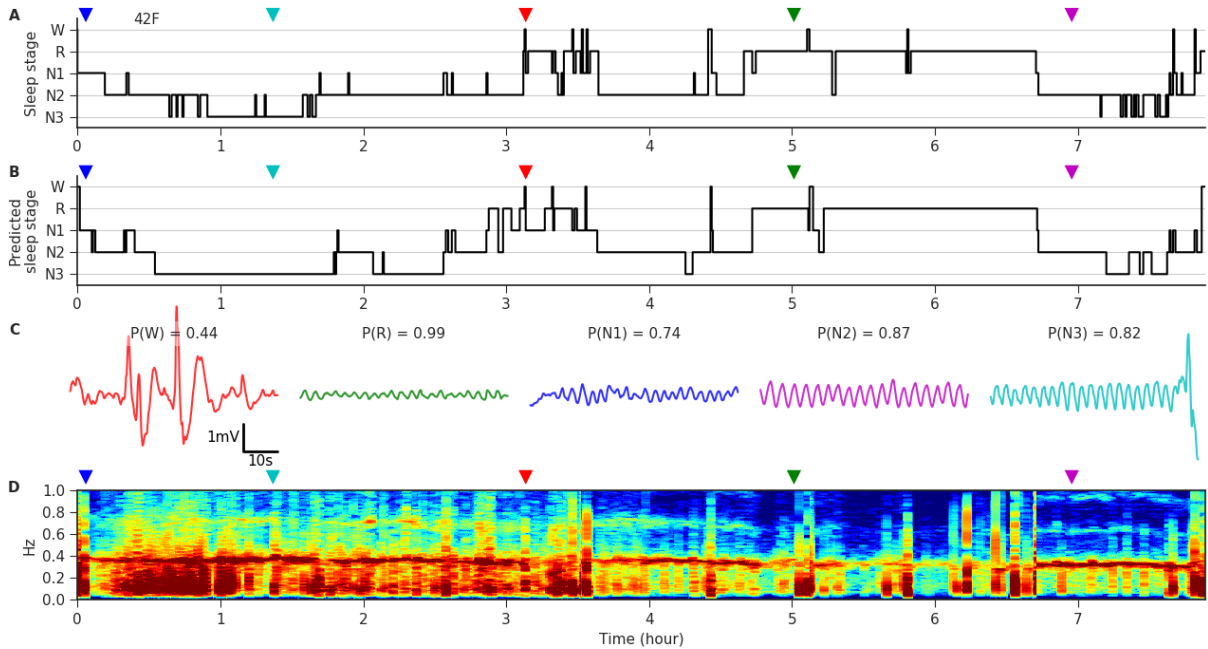

**Figure 7**. Similar to Figure 6, showing an example 42-year female using CHEST respiration as input. The scaling of the signals in Panel C is the same as in Figure 6, but amplitude of these example signal itself happens to be smaller. It is possible that other epochs have larger or similar amplitude compared to Figure 6.

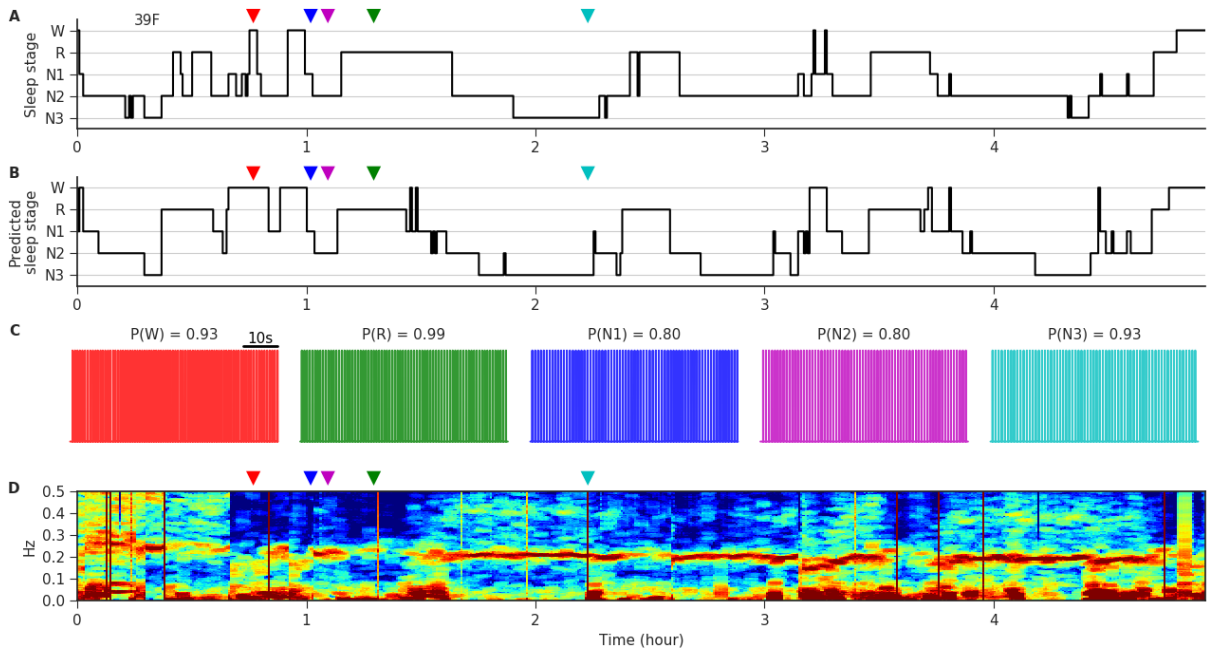

**Figure 8**. Similar to Figure 6, showing an example 39-year female using the R peaks from ECG as input. Panel C shows the R peaks represented as a binary sequence (see Methods). Panel D shows the spectrogram of the R peak intervals [21].

## Discussion

We hypothesize that it is possible to accurately stage sleep based on the electrocardiogram (ECG) and respiratory signals using deep learning. Our key findings are: 1) ECG and respiratory signals contain substantial information about sleep stages; 2) reduced staging accuracy is associated with older age and/or more severe sleep apnea, although the networks still perform well on signals from patients with advanced age and high AHI; 3) using deep learning, the staging performance is robust for a wide range of typical sleep disorders like OSA and PLMS, and commonly used medications; 4) collapsing certain stages of sleep/wake (e.g., N1 with wake and N2 with N3) results in greater staging agreement.

Previous studies comparable to ours are summarized in Table 4. Some prior studies have also sought to stage sleep from ECG and respiration using deep neural networks. However, these studies suffered from small sample sizes and limiting generalizability. Only one prior study used more than 100 participants for training and evaluation. The large sample size of training and testing sets in the present study provides more robust results compared to prior literature, and is expected to increase generalizability when applied to heterogeneous / external populations.

**Table 4**. Related work in the literature.

| Author (year) | Dataset size | Performance (κ is Cohen's kappa) | Type of Signal | Features |
|---|---|---|---|---|
| Sady et al. 2013 [12] | 13 participants | 3 stages (W, NREM, REM): accuracy=78%<br>5 stages (W, N1, N2, N3, REM): accuracy=62% | Photo-Plethysmogram, hemoglobin oxygen saturation, pneumotachograph | Heartbeat interval, time domain respiratory signals. |
| Long et al. 2014 [22] | 48 participants | 3 stages (W, NREM, REM): κ=0.48<br>4 stages (W, light sleep, deep sleep, REM): κ =0.41 | Respiratory effort | Time domain, dissimilarity measure |
| Fonseca et al. 2015 [23] | 48 participants | 4 stages (W, light sleep, deep sleep, REM): κ=0.49, accuracy=69%<br>3 stages (W, NREM, REM): κ=0.56<br>Accuracy=80% | ECG + Respiratory inductance plethysmography | Time and frequency domain, nonlinear |
| Zhao et al. 2017 [24] | 25 participants, 100 nights | 4 stages (W, N1+N2, N3, REM): κ=0.70, accuracy=79.8% | Radio frequency signal reflected off body (heartbeat, respiration) | Radio frequency spectrogram |
| Zhang et al. 2017 [25] | 37,000 epochs | 5 stages (W, N1, N2, N3, REM): Precision = 53.9%<br>Recall = 56.0%<br>F1 score = 53.2% | Heart rate derived from a wearable device | Frequency domain (DCT) |

| Radha et al. 2018 [26] | ECG: 352 participants, PPG: 60 participants | ECG: 6 stages (W, S1, S2, S3, S4, REM), κ=0.61 and accuracy=76.30% PPG: 5 stages (W, N1, N2, N3, REM), κ=0.63 and accuracy=74.65% | ECG and PPG | Selected features from time and frequency domain |
|---|---|---|---|---|

Sleep staging based on ECG and respiration has lower performance compared to using EEG. We previously performed EEG-based sleep staging with a deep neural network trained on data from the same set of patients used in the present work. This achieves performance similar to human inter-rater agreement [17]. This is not surprising since sleep technicians stage sleep mainly using EEG based on the AASM guideline.

The improvement in staging performance when collapsing certain stages of sleep into super-stages may reflect information regarding the true biology of sleep states. N1 is an unstable transitional state with low probability and non-distinct EEG features. About half of sleep is N2, and can show both stable and unstable characteristics, such as cyclic alternating pattern, apneic, or stable breathing in patients with sleep apnea. Different methods to characterize sleep depth and quality are available, and it will be important in future work to investigate whether further parsing of NREM sleep is meaningful using machine learning combined with methods such as the Odds Ratio Product of NREM sleep depth [27] or ECG-cardiopulmonary coupling [8].

The mild degradation of performance with age is not surprising when using conventional sleep stages as the ground truth. The reduction of N3 with age (mainly in males) is not accompanied by equal and simultaneous reductions in stable N2 – thus, older individuals with equally reduced N3 may have very different N2 quality. By contrast, stable N2 and N3 may have very similar or identical cardiorespiratory signatures, making it difficult or impossible for deep learning models to reliably distinguish them. Thus "errors" in discriminating these stages may reflect that EEG-based annotation in the reference standard is somewhat orthogonal to autonomic fluctuations.

Estimation of sleep states from cardiac and respiratory signals can simplify sleep tracking in health and disease, especially in environments like an intensive care unit (ICU) or hospitalized patients in general, when the model is trained with enough ICU patients who receive various heart rate or blood pressure medications.

Limitations of our analysis are as follows. 1) Our dataset includes only adults, and generalizability to the pediatric group will require additional study. 2) The 30-second epoch-based scoring of sleep limits the fine-grained analysis of sleep stages. This is especially true when sleep fragmenting conditions are present, where a given 30-second epoch may have features of multiple states. Moreover, boundary zones may be amplified, such as transitions between wake-REM and NREM in the presence of sleep apnea in REM sleep. Such periods will introduce "error" in machine learning analyses, though these are biological features of sleep fragmentation rather than measurement or characterization error, such as arousal, apnea, or limb movement. 3) Due to the "black-box" nature of deep neural networks, there is limited insight into what the networks use as key features. Future work to interpret what the networks have learned (beyond Figure 6, Figure 7, Figure 8, and Figure S4-S12 in the supplementary material) is needed. (4) 1-fold validation (single training-validation-testing split) is used. Although this is the common practice in large datasets, it is nevertheless less biased to use cross-validation on multiple folds.

In conclusion, utilizing a large-scale dataset consisting of 8,682 PSGs, we have developed a set of deep neural networks to classify sleep stages from ECG and/or respiration. ECG and respiratory effort provide substantial information about sleep stages. The best staging performance is obtained using both ECG and abdominal respiration. Staging performance depends to some extent on age, apnea-hypopnea index, and sleep study type.

# Acknowledgments

We gratefully acknowledge expert technical support from the Clinical Data Animation Center (CDAC) at Massachusetts General Hospital.

# Disclosure Statement

RJT reports 1) Patent, license and royalties from MyCardio, LLC, for an ECG-based method to phenotype sleep quality and sleep apnea; 2) GLG consulting for general sleep medicine; 3) Intellectual Property (patent) for a device using $CO_2$ for central / complex sleep apnea. BG declares that the work was done while at Massachusetts General Hospital. He is currently a full time employee at Novartis Institutes of Biomedical Research with a role of Data Scientist.

# References

1.	Silber MH, Ancoli-Israel S, Bonnet MH, et al. The visual scoring of sleep in adults. J Clin Sleep Med. 2007; 3 (2): 121-131.
2.	Chervin RD, Shelgikar AV, Burns JW. Respiratory cycle-related EEG changes: response to CPAP. Sleep. 2012; 35 (2): 203-209.
3.	Niizeki K, Saitoh T. Association Between Phase Coupling of Respiratory Sinus Arrhythmia and Slow Wave Brain Activity During Sleep. Front Physiol. 2018; 9: 1338.
4.	Penzel T, Kantelhardt JW, Bartsch RP, et al. Modulations of Heart Rate, ECG, and Cardio-Respiratory Coupling Observed in Polysomnography. Front Physiol. 2016; 7: 460.
5.	Thomas RJ, Mietus JE, Peng CK, et al. Relationship between delta power and the electrocardiogram-derived cardiopulmonary spectrogram: possible implications for assessing the effectiveness of sleep. Sleep Med. 2014; 15 (1): 125-131.
6.	Lockmann AL, Laplagne DA, Leao RN, Tort AB. A Respiration-Coupled Rhythm in the Rat Hippocampus Independent of Theta and Slow Oscillations. J Neurosci. 2016; 36 (19): 5338-5352.
7.	Iellamo F, Placidi F, Marciani MG, et al. Baroreflex buffering of sympathetic activation during sleep: evidence from autonomic assessment of sleep macroarchitecture and microarchitecture. Hypertension. 2004; 43 (4): 814-819.

8. Thomas RJ, Mietus JE, Peng CK, Goldberger AL. An electrocardiogram-based technique to assess cardiopulmonary coupling during sleep. Sleep. 2005; 28 (9): 1151-1161.
9. Sei H. Blood pressure surges in REM sleep: A mini review. Pathophysiology. 2012; 19 (4): 233-241.
10. Thomas RJ, Wood C, Bianchi MT. Cardiopulmonary coupling spectrogram as an ambulatory clinical biomarker of sleep stability and quality in health, sleep apnea and insomnia. Sleep. 2017.
11. Bianchi MT. Sleep devices: wearables and nearables, informational and interventional, consumer and clinical. Metabolism. 2018; 84: 99-108.
12. Sady CC, Freitas US, Portmann A, Muir JF, Letellier C, Aguirre LA. Automatic sleep staging from ventilator signals in non-invasive ventilation. Comput Biol Med. 2013; 43 (7): 833-839.
13. Migliorini M, Bianchi AM, Nistico D, et al. Automatic sleep staging based on ballistocardiographic signals recorded through bed sensors. Conf Proc IEEE Eng Med Biol Soc. 2010; 2010: 3273-3276.
14. Tal A, Shinar Z, Shaki D, Codish S, Goldbart A. Validation of Contact-Free Sleep Monitoring Device with Comparison to Polysomnography. J Clin Sleep Med. 2017; 13 (3): 517-522.
15. Zaffaroni A, Doheny EP, Gahan L, et al. Non-Contact Estimation of Sleep Staging. 2018; Singapore.
16. Watson PL, Pandharipande P, Gehlbach BK, et al. Atypical sleep in ventilated patients: empirical electroencephalography findings and the path toward revised ICU sleep scoring criteria. Crit Care Med. 2013; 41 (8): 1958-1967.
17. Biswal S, Sun H, Goparaju B, Westover MB, Sun J, Bianchi MT. Expert-level sleep scoring with deep neural networks. Journal of the American Medical Informatics Association. 2018; 25 (12): 1643-1650.
18. Pan J, Tompkins WJ. A real-time QRS detection algorithm. IEEE Trans Biomed Eng. 1985; 32 (3): 230-236.
19. Rebergen DJ, Nagaraj SB, Rosenthal ES, Bianchi MT, van Putten MJ, Westover MB. ADARRI: a novel method to detect spurious R-peaks in the electrocardiogram for heart rate variability analysis in the intensive care unit. Journal of clinical monitoring and computing. 2018; 32 (1): 53-61.
20. Hannun AY, Rajpurkar P, Haghpanahi M, et al. Cardiologist-level arrhythmia detection and classification in ambulatory electrocardiograms using a deep neural network. Nature medicine. 2019; 25 (1): 65.
21. van Gent PaF, Haneen and Nes, Nicole and Arem, B. Heart Rate Analysis for Human Factors: Development and Validation of an Open Source Toolkit for Noisy Naturalistic Heart Rate Data. In: proceedings from the the 6th HUMANIST Conference; 2018; the Netherlands.
22. Long X, Yang J, Weysen T, et al. Measuring dissimilarity between respiratory effort signals based on uniform scaling for sleep staging. Physiological measurement. 2014; 35 (12): 2529.
23. Fonseca P, Long X, Radha M, Haakma R, Aarts RM, Rolink J. Sleep stage classification with ECG and respiratory effort. Physiological measurement. 2015; 36 (10): 2027.
24. Zhao M, Yue S, Katabi D, Jaakkola TS, Bianchi MT. Learning sleep stages from radio signals: A conditional adversarial architecture. In: proceedings from the Proceedings of the 34th International Conference on Machine Learning-Volume 70; 2017.
25. Zhang X, Kou W, Eric I, et al. Sleep stage classification based on multi-level feature learning and recurrent neural networks via wearable device. Computers in biology and medicine. 2018; 103: 71-81.
26. Radha M, Fonseca P, Ross M, Cerny A, Anderer P, Aarts RM. LSTM knowledge transfer for HRV-based sleep staging. arXiv preprint arXiv:180906221. 2018.
27. Younes M, Ostrowski M, Soiferman M, et al. Odds ratio product of sleep EEG as a continuous measure of sleep state. Sleep. 2015; 38 (4): 641-654.

**Supplementary Figures**

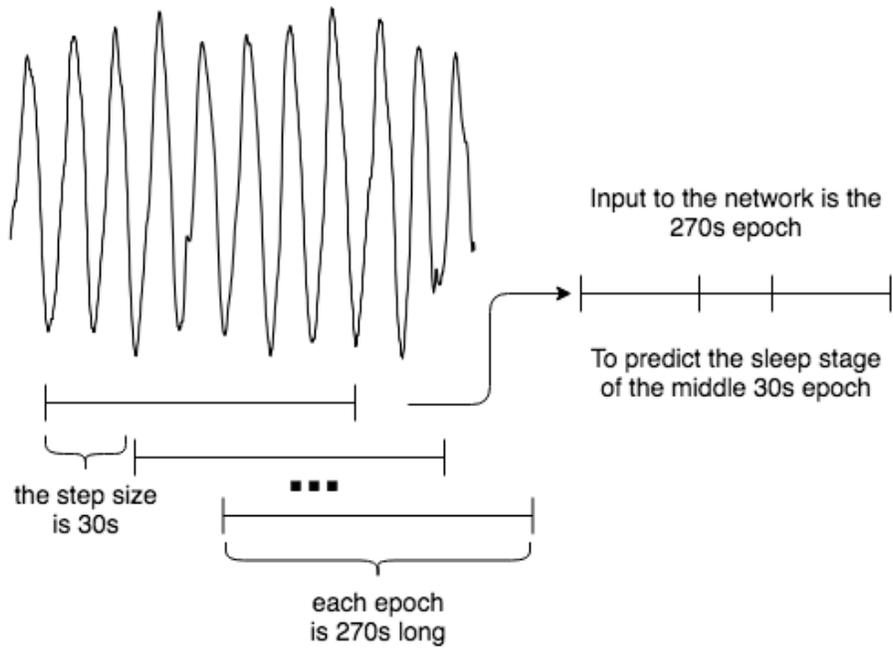

**Figure S1**. Illustration of signal segmentation.

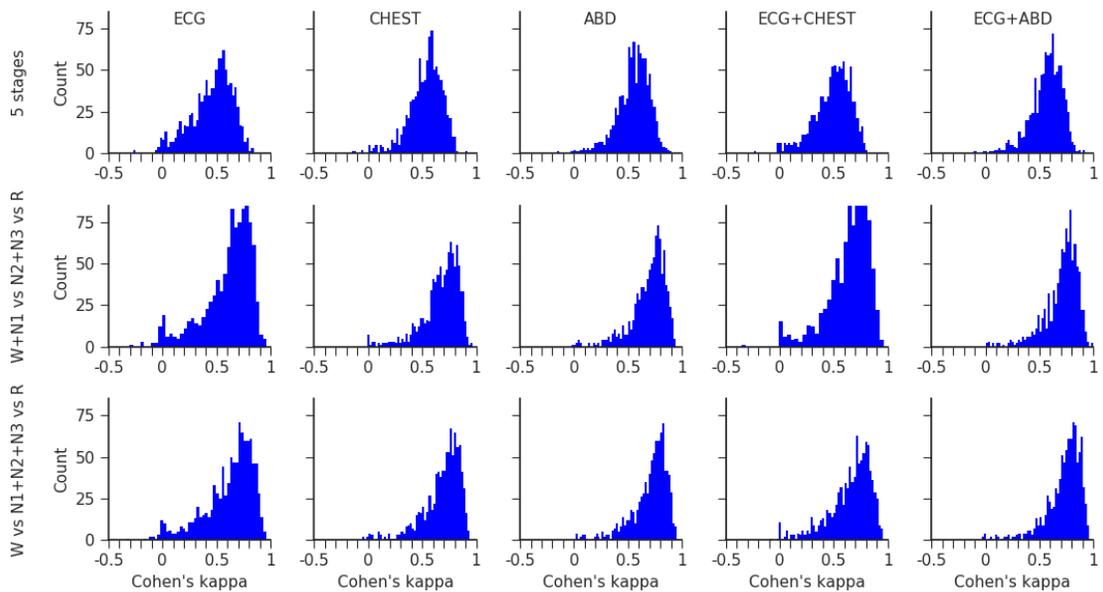

**Figure S2.** The histogram of Cohen's kappa of individual PSGs for different input signals and different combinations of sleep stages.

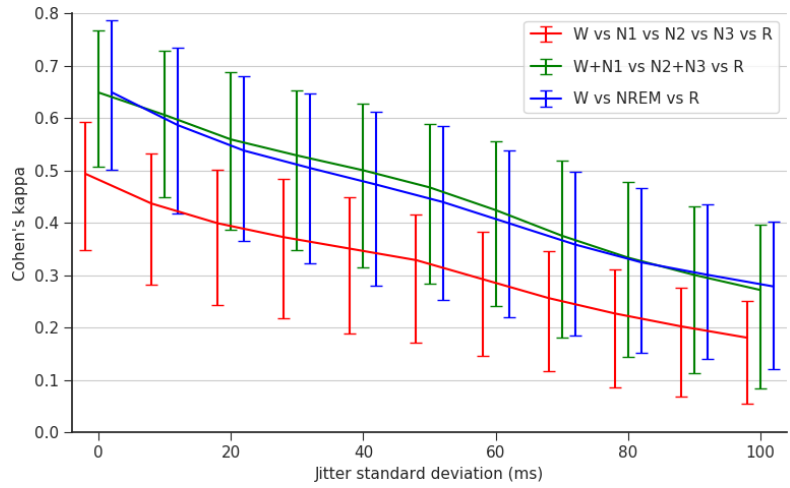

**Figure S3**. Cohen's kappa when applying zero-mean Gaussian jitter to the ECG R peaks.

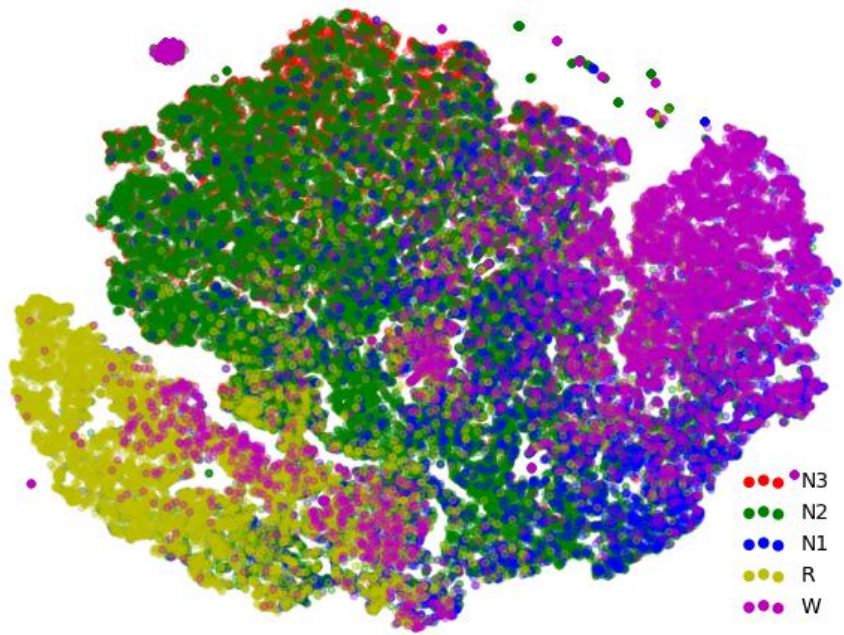

**Figure S4**. tSNE visualization of the last layer activation of the deep network that takes ECG as the input. Each point in the figure is a 270-second epoch.

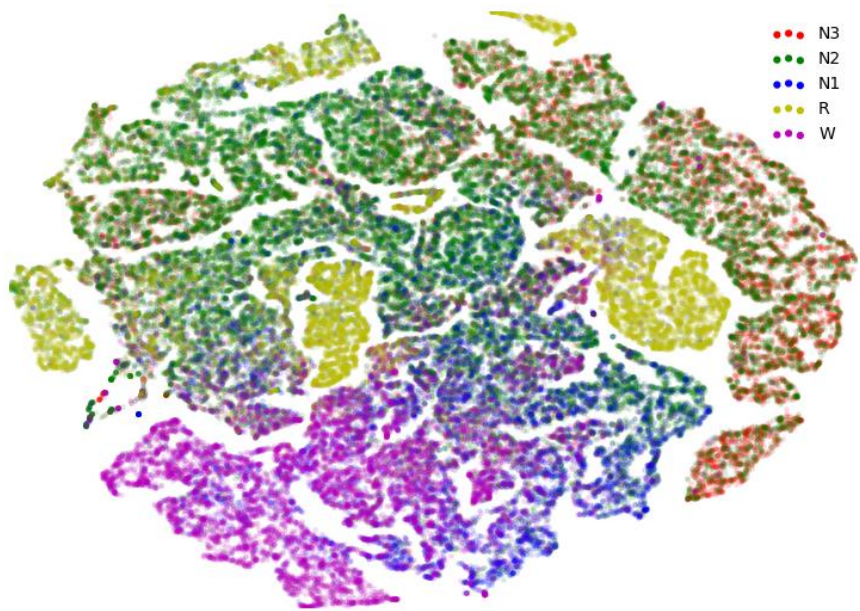

**Figure S5**. tSNE visualization of the last layer activation of the deep network that takes chest repiration signal as the input. Each point in the figure is a 270-second epoch.

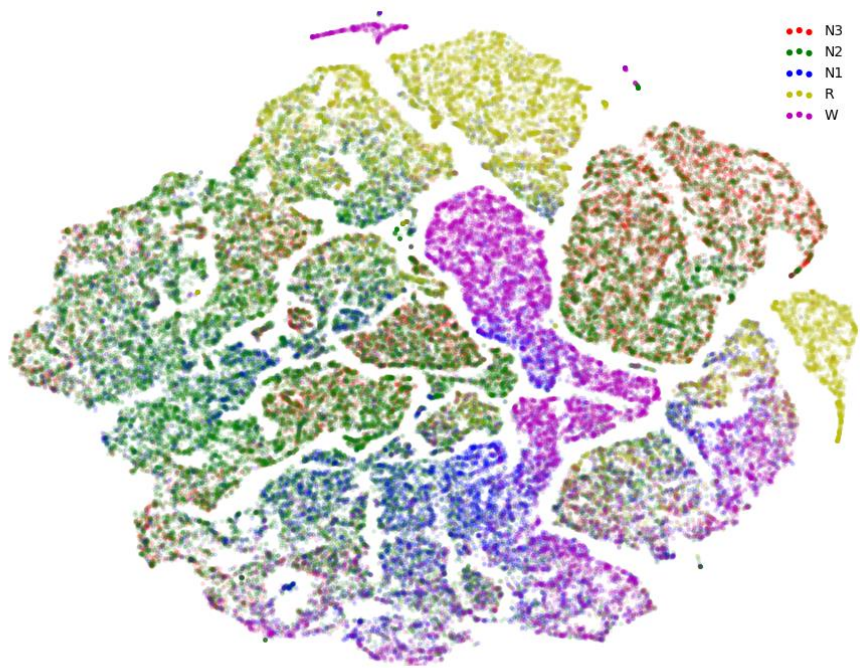

**Figure S6**. tSNE visualization of the last layer activation of the deep network that takes abdominal repiration as the input. Each point in the figure is a 270-second epoch.

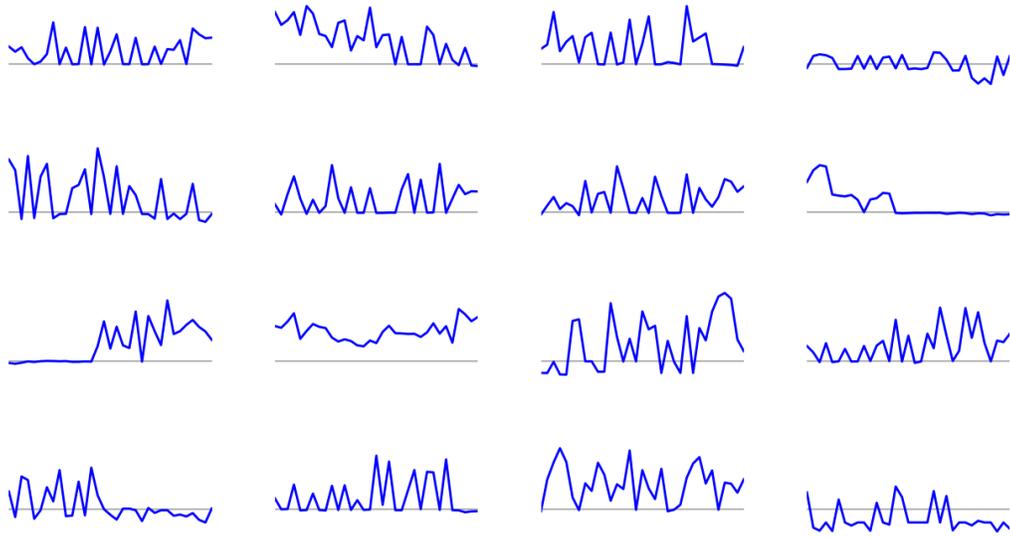

**Figure S7**. The kernels of the first convolution layer in the deep network that takes ECG as the input.

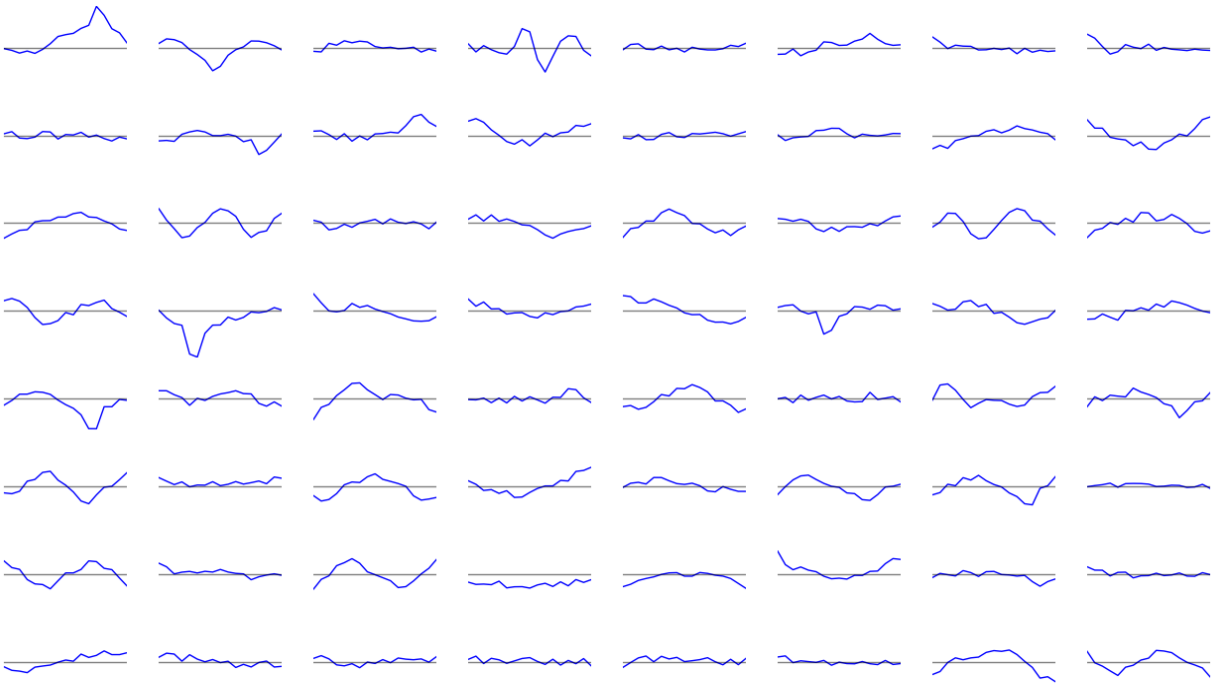

**Figure S8**. The kernels of the first convolution layer in the deep network that takes chest respiration as the input.

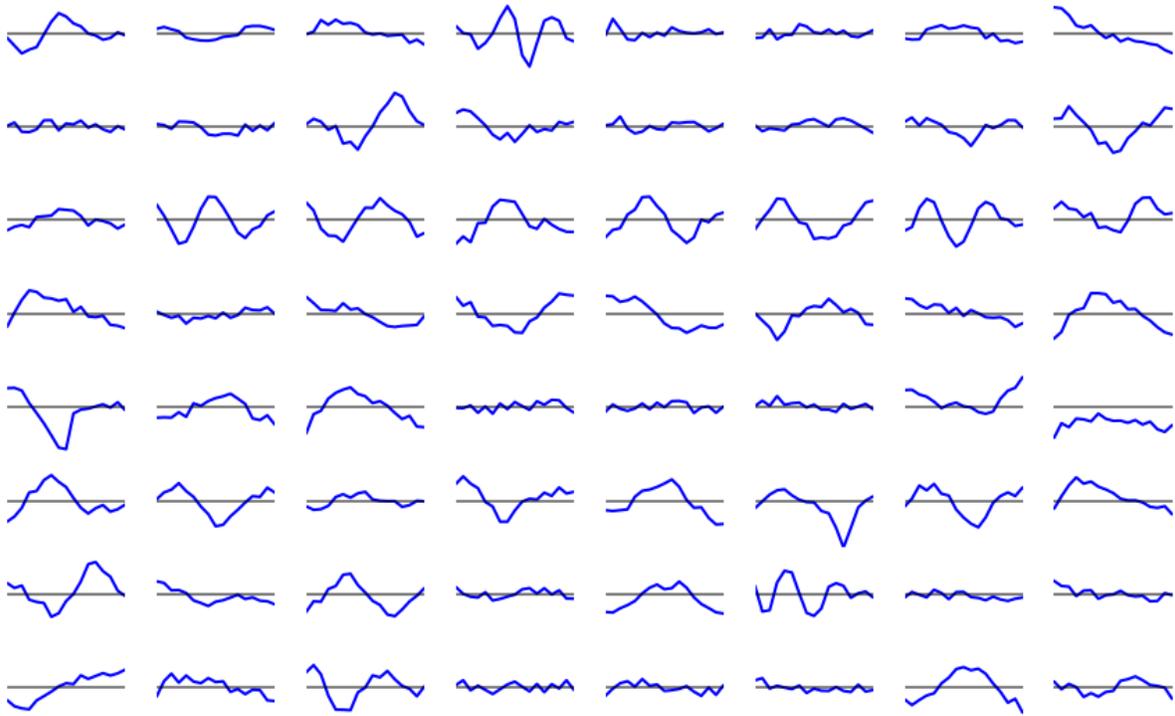

**Figure S9**. The kernels of the first convolution layer in the deep network that takes abdominal respiration as the input.

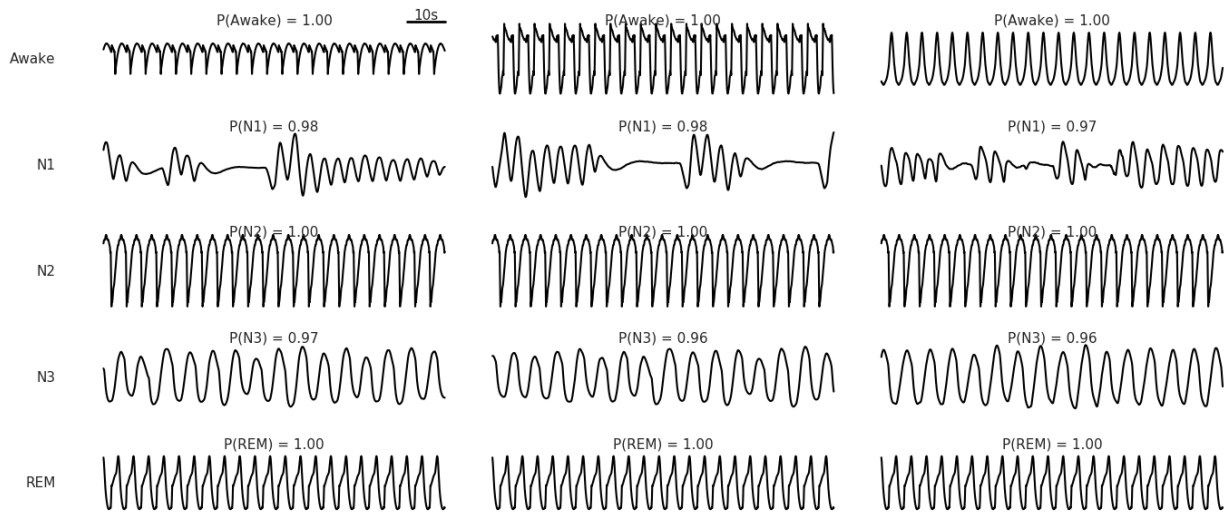

**Figure S10**. Examples of ABD signals for different sleep stages. These examples are selected based on having high probability according to the deep neural network in each of the 5 sleep stages. The signals are not necessarily from the same recording.

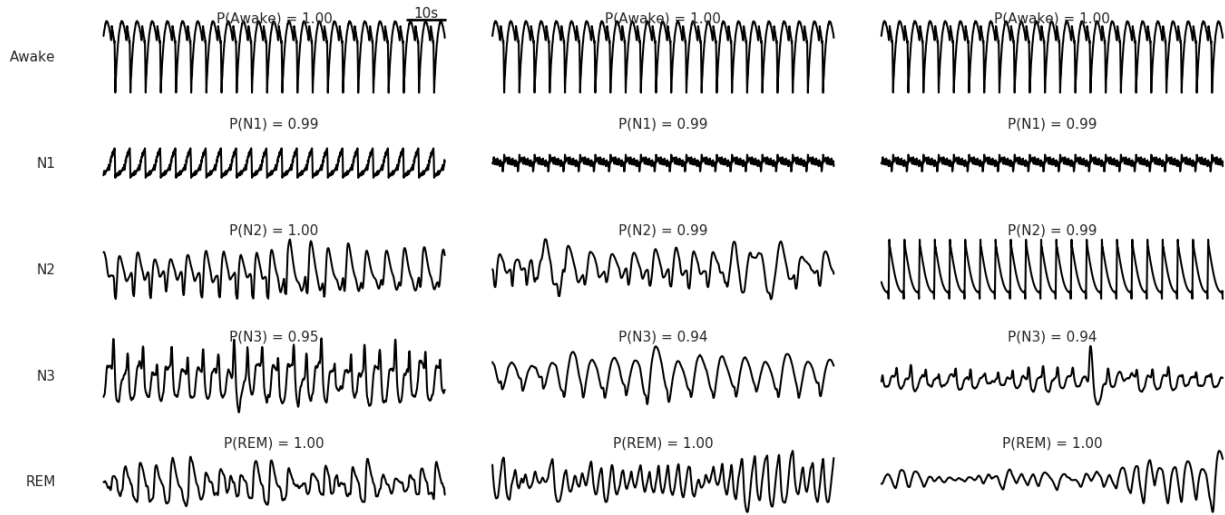

**Figure S11**. Examples of CHEST signals in different sleep stages, selected based on having high probability by the deep neural network. The signals are not necessarily from the same recording.

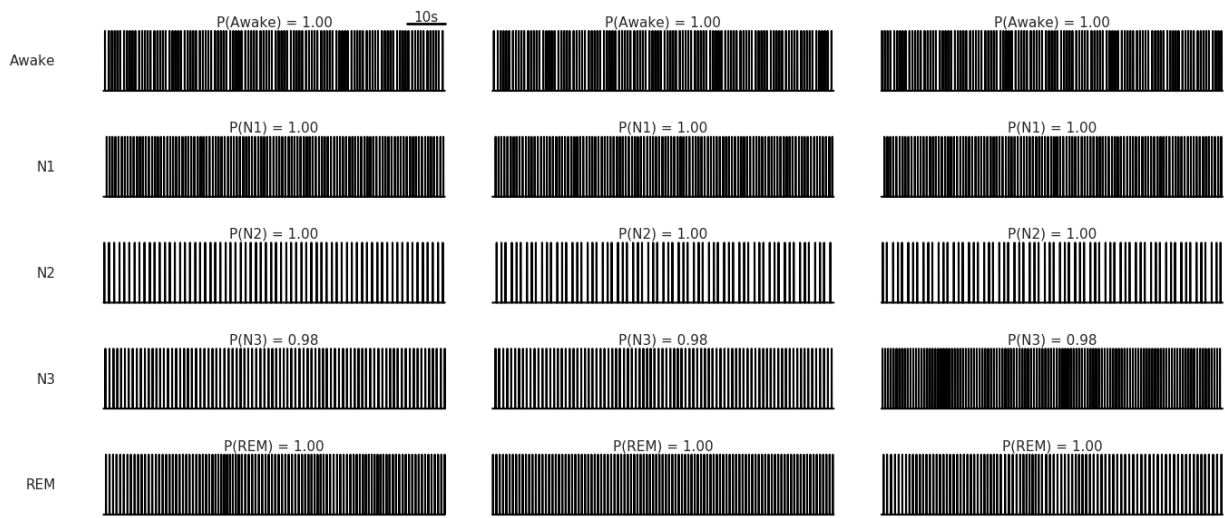

**Figure S12**. Example ECG R peaks in different sleep stages, selected based on having high probability by the deep neural network. The signals are not necessarily from the same recording.

**Supplementary Tables**

Table S1. Cohen's kappa in different group in the 1000-PSG testing set using ECG as input.

| Category | Group | 5 stages | 3 stages W+N1 vs. N2+N3 vs. R | 3 stages W vs. NR vs. R |
|---|---|---|---|---|
| Age | Young: 18 ≤ Age < 40 | 0.54 | 0.715 | 0.717 |
| | Middle: 40 ≤ Age < 60 | 0.514 | 0.672 | 0.685 |
| | Elderly: Age ≥ 60 | 0.426 | 0.57 | 0.564 |
| Sex | Male | 0.483 | 0.636 | 0.644 |
| | Female | 0.51 | 0.673 | 0.67 |
| BMI (kg/m$^2$) | Normal: 18.5 ≤ BMI < 25 | 0.536 | 0.678 | 0.672 |
| | Overweight: BMI ≥ 25 | 0.49 | 0.65 | 0.654 |
| Type of Test | Diagnostic | 0.508 | 0.667 | 0.667 |
| | All Night CPAP | 0.501 | 0.67 | 0.663 |
| | Split Night | 0.462 | 0.601 | 0.622 |
| AHI (per hour) | Normal: AHI < 5 | 0.519 | 0.699 | 0.701 |
| | Mild: 5 ≤ AHI < 15 | 0.499 | 0.655 | 0.639 |
| | Moderate: 15 ≤ AHI < 30 | 0.471 | 0.615 | 0.621 |
| | Severe: AHI ≥ 30 | 0.427 | 0.532 | 0.603 |
| Periodic limb movement (per hour) | Normal: PLMI < 5 | 0.497 | 0.656 | 0.676 |
| | Mild: 5 ≤ PLMI < 15 | 0.512 | 0.667 | 0.669 |
| | Moderate: 15 ≤ PLMI < 30 | 0.502 | 0.664 | 0.649 |
| | Severe: PLMI ≥ 30 | 0.459 | 0.609 | 0.611 |
| Medication | Antidepressant | 0.489 | 0.651 | 0.652 |
| | Benzodiazepine | 0.501 | 0.674 | 0.662 |
| | Diabetic | 0.501 | 0.673 | 0.668 |
| | Herbal | 0.484 | 0.65 | 0.639 |
| | Hypertension | 0.495 | 0.649 | 0.655 |
| | Neuroleptic | 0.483 | 0.664 | 0.674 |
| | Opiate | 0.476 | 0.624 | 0.626 |
| | Neuroactive | 0.5 | 0.648 | 0.661 |
| | Systemic | 0.496 | 0.656 | 0.663 |
| | RLS/PLMS | 0.515 | 0.647 | 0.657 |
| | Sleeping | 0.499 | 0.669 | 0.66 |
| | Stimulant | 0.507 | 0.671 | 0.646 |
| | Z-drug | 0.526 | 0.692 | 0.682 |

**Table S2**. Cohen's kappa in different group in the 1000-PSG testing set using CHEST as input.

| Category | Group | 5 stages | 3 stages | |
| --- | --- | --- | --- | --- |
| | | | W+N1 vs. N2+N3 vs. R | W vs. NR vs. R |
| Age | Young: 18 ≤ Age < 40 | 0.581 | 0.728 | 0.728 |
| | Middle: 40 ≤ Age < 60 | 0.569 | 0.709 | 0.713 |
| | Old: Age ≥ 60 | 0.546 | 0.691 | 0.689 |
| Sex | Male | 0.558 | 0.701 | 0.704 |
| | Female | 0.573 | 0.719 | 0.716 |
| BMI (kg/m$^2$) | Normal: 18.5 ≤ BMI < 25 | 0.554 | 0.697 | 0.692 |
| | Overweight: BMI ≥ 25 | 0.567 | 0.71 | 0.711 |
| Type of Test | Diagnostic | 0.57 | 0.714 | 0.713 |
| | All Night CPAP | 0.564 | 0.719 | 0.702 |
| | Split Night | 0.553 | 0.685 | 0.707 |
| AHI (per hour) | Normal: AHI < 5 | 0.576 | 0.734 | 0.725 |
| | Mild: 5 ≤ AHI < 15 | 0.572 | 0.714 | 0.708 |
| | Moderate: 15 ≤ AHI < 30 | 0.555 | 0.688 | 0.696 |
| | Severe: AHI ≥ 30 | 0.517 | 0.628 | 0.68 |
| Periodic limb movement (per hour) | Normal: PLMI < 5 | 0.556 | 0.698 | 0.71 |
| | Mild: 5 ≤ PLMI < 15 | 0.576 | 0.715 | 0.709 |
| | Moderate: 15 ≤ PLMI < 30 | 0.576 | 0.72 | 0.716 |
| | Severe: PLMI ≥ 30 | 0.551 | 0.698 | 0.699 |
| Medication | Antidepressant | 0.573 | 0.72 | 0.718 |
| | Benzodiazepine | 0.572 | 0.714 | 0.706 |
| | Diabetic | 0.561 | 0.708 | 0.707 |
| | Herbal | 0.586 | 0.728 | 0.713 |
| | Hypertension | 0.565 | 0.704 | 0.709 |
| | Neuroleptic | 0.532 | 0.691 | 0.689 |
| | Opiate | 0.576 | 0.709 | 0.704 |
| | Neuroactive | 0.573 | 0.713 | 0.72 |
| | Systemic | 0.564 | 0.706 | 0.71 |
| | RLS/PLMS | 0.58 | 0.725 | 0.726 |
| | Sleeping | 0.572 | 0.719 | 0.713 |
| | Stimulant | 0.581 | 0.722 | 0.7 |
| | Z-drug | 0.581 | 0.72 | 0.709 |

**Table S3**. Cohen's kappa in different group in the 1000-PSG testing set using ABD as input.

| Category | Group | 5 stages | 3 stages | |
|---|---|---|---|---|
| | | | W+N1 vs. N2+N3 vs. R | W vs. NR vs. R |
| Age | Young: 18 ≤ Age < 40 | 0.584 | 0.737 | 0.753 |
| | Middle: 40 ≤ Age < 60 | 0.586 | 0.728 | 0.746 |
| | Old: Age ≥ 60 | 0.551 | 0.698 | 0.705 |
| Sex | Male | 0.574 | 0.719 | 0.737 |
| | Female | 0.577 | 0.725 | 0.732 |
| BMI (kg/m$^2$) | Normal: 18.5 ≤ BMI < 25 | 0.552 | 0.684 | 0.679 |
| | Overweight: BMI ≥ 25 | 0.579 | 0.727 | 0.743 |
| Type of Test | Diagnostic | 0.584 | 0.724 | 0.733 |
| | All Night CPAP | 0.578 | 0.738 | 0.734 |
| | Split Night | 0.555 | 0.698 | 0.737 |
| AHI (per hour) | Normal: AHI < 5 | 0.584 | 0.742 | 0.741 |
| | Mild: 5 ≤ AHI < 15 | 0.585 | 0.734 | 0.739 |
| | Moderate: 15 ≤ AHI < 30 | 0.563 | 0.696 | 0.723 |
| | Severe: AHI ≥ 30 | 0.533 | 0.651 | 0.725 |
| Periodic limb movement (per hour) | Normal: PLMI < 5 | 0.574 | 0.726 | 0.748 |
| | Mild: 5 ≤ PLMI < 15 | 0.583 | 0.73 | 0.738 |
| | Moderate: 15 ≤ PLMI < 30 | 0.579 | 0.714 | 0.728 |
| | Severe: PLMI ≥ 30 | 0.559 | 0.7 | 0.714 |
| Medication | Antidepressant | 0.574 | 0.731 | 0.743 |
| | Benzodiazepine | 0.595 | 0.738 | 0.749 |
| | Diabetic | 0.574 | 0.736 | 0.74 |
| | Herbal | 0.605 | 0.741 | 0.75 |
| | Hypertension | 0.576 | 0.719 | 0.738 |
| | Neuroleptic | 0.527 | 0.695 | 0.715 |
| | Opiate | 0.585 | 0.722 | 0.74 |
| | Neuroactive | 0.581 | 0.729 | 0.743 |
| | Systemic | 0.578 | 0.726 | 0.744 |
| | RLS/PLMS | 0.58 | 0.739 | 0.74 |
| | Sleeping | 0.589 | 0.737 | 0.745 |
| | Stimulant | 0.604 | 0.745 | 0.769 |
| | Z-drug | 0.602 | 0.75 | 0.756 |

**Table S4**. Cohen's kappa in different group in the 1000-PSG testing set using ECG + CHEST as input.

| Category | Group | 5 stages | 3 stages ||
| --- | --- | --- | --- | --- |
| | | | W+N1 vs. N2+N3 vs. R | W vs. NR vs. R |
| Age | Young: 18 ≤ Age < 40 | 0.539 | 0.691 | 0.708 |
| | Middle: 40 ≤ Age < 60 | 0.506 | 0.645 | 0.657 |
| | Old: Age ≥ 60 | 0.468 | 0.603 | 0.621 |
| Sex | Male | 0.492 | 0.633 | 0.647 |
| | Female | 0.518 | 0.659 | 0.673 |
| BMI (kg/m$^2$) | Normal: 18.5 ≤ BMI < 25 | 0.54 | 0.672 | 0.682 |
| | Overweight: BMI ≥ 25 | 0.499 | 0.642 | 0.656 |
| Type of Test | Diagnostic | 0.518 | 0.661 | 0.674 |
| | All Night CPAP | 0.504 | 0.652 | 0.644 |
| | Split Night | 0.472 | 0.602 | 0.642 |
| AHI (per hour) | Normal: AHI < 5 | 0.521 | 0.677 | 0.683 |
| | Mild: 5 ≤ AHI < 15 | 0.518 | 0.657 | 0.651 |
| | Moderate: 15 ≤ AHI < 30 | 0.472 | 0.604 | 0.637 |
| | Severe: AHI ≥ 30 | 0.451 | 0.554 | 0.627 |
| Periodic limb movement (per hour) | Normal: PLMI < 5 | 0.5 | 0.643 | 0.668 |
| | Mild: 5 ≤ PLMI < 15 | 0.52 | 0.655 | 0.661 |
| | Moderate: 15 ≤ PLMI < 30 | 0.511 | 0.657 | 0.661 |
| | Severe: PLMI ≥ 30 | 0.475 | 0.611 | 0.632 |
| Medication | Antidepressant | 0.508 | 0.66 | 0.664 |
| | Benzodiazepine | 0.519 | 0.663 | 0.673 |
| | Diabetic | 0.488 | 0.658 | 0.658 |
| | Herbal | 0.547 | 0.687 | 0.719 |
| | Hypertension | 0.49 | 0.631 | 0.641 |
| | Neuroleptic | 0.462 | 0.633 | 0.679 |
| | Opiate | 0.499 | 0.613 | 0.644 |
| | Neuroactive | 0.504 | 0.648 | 0.667 |
| | Systemic | 0.499 | 0.643 | 0.656 |
| | RLS/PLMS | 0.516 | 0.671 | 0.681 |
| | Sleeping | 0.517 | 0.667 | 0.679 |
| | Stimulant | 0.539 | 0.677 | 0.683 |
| | Z-drug | 0.547 | 0.684 | 0.703 |